\documentstyle[11pt,newpasp,twoside,epsf]{article}
\def\edcomment#1{\iffalse\marginpar{\raggedright\sl#1\/}\else\relax\fi}
\reversemarginpar

\begin{document}
\title{Molecular Gas Chemistry in Starbursts and AGN}
 \author{S. Garc\'{\i}a-Burillo, A. Fuente, A. Usero}
\affil{Observatorio Astron\'omico Nacional (OAN), C/Alfonso XII, 3, 28014-Madrid, SPAIN}
\author{J. Mart\'{\i}n-Pintado}
\affil{Instituto de Estructura de la Materia (IEM), CSIC, Serrano 121, 28006-Madrid, SPAIN }

\begin{abstract}
We present the main results of an extragalactic survey aiming to study the chemistry of molecular
gas in a limited sample of starburst galaxies (SB) and AGN hosts. Observations have been carried
out with the IRAM 30m telescope and the Plateau de Bure Interferometer (PdBI). The high
resolution/sensitivity of the PdBI has made possible to obtain high quality images of the galaxies
using specific molecular gas tracers of Shock Chemistry, Photon Dominated Regions (PDR) and X-ray
Dominated Regions (XDR). The occurrence of large-scale shocks and the propagation of PDR chemistry
in starbursts can be studied. We also discuss the onset of XDR chemistry in AGN.
\end{abstract}

\section{Evolution of Molecular Gas Chemistry in SB and AGN}

Multi-wavelength based evidence indicates that the properties of molecular gas in SB and
AGN hosts differ from that of quiescent star forming disks. The spectacular energies injected in the
gas reservoirs of SB and AGN coming as strong radiation fields (UV, X-rays,...), powerful winds and
jets can create a particularly harsh environment for ISM. A complete understanding of the
physical/chemical evolution of molecular gas in `active' galaxies requires the use of specific
tracers of the relevant energetic phenomena that are at work all the way along the starburst
sequence: large-scale shocks, strong UV-fields and nuclear X-ray irradiation. We have used the
high-resolution/sensitivity of the IRAM 30m telescope and the PdBI to study a limited sample of
prototypical SB and AGN.

\subsection{Unveiling Shock Chemistry in SB}

We have carried out a 30m multi-transition survey searching for the thermal emission of silicon
monoxide (SiO) in a dozen prototypical SB galaxies including NGC\,253, M\,82 and NGC\,1068. Studies
of galactic clouds point out to SiO as a privileged tracer of shocks in the molecular gas phase.
Shocks can significantly increase SiO fractional abundances to X(SiO)$\sim$10$^{-8}$. Our survey
shows that SiO emission is widespread in the circumnuclear disks (CND) of SB on scales ranging from
100\,pc to 700\,pc. The estimated SiO abundances vary within the sample from  X(SiO)$\sim$10$^{-9}$
in NGC\,253 to 1/50 of this value in M\,82. Physical parameters of SiO clouds are also very
different, suggesting different scenarios for shock chemistry.

The first PdBI SiO maps obtained in NGC\,253 and M\,82 have given invaluable insight into the
different mechanisms driving large-scale shocks in the molecular gas disks of SB 
(Garc\'{\i}a-Burillo et al. 2000, 2001). While SiO emission is detected mainly in a 700\,pc-diameter
CND in NGC\,253 (Garc\'{\i}a-Burillo et al. 2000), the emission of SiO extends noticeably out of
the galaxy plane in M\,82, tracing the disk-halo interface where episodes of mass injection from
the disk are building up the gaseous halo (Garc\'{\i}a-Burillo et al. 2001; see Fig. 1). 
Large-scale shocks are driven by massive star formation and bar density waves inside the disk of
NGC\,253. In M\,82, however, shocked molecular gas appears forming a 500\,pc chimney and a giant
supershell. The strikingly different distributions and average fractional abundances of SiO in
NGC\,253 and M\,82 are suggestive of an evolutionary link between these SB; the latter pictures the
M\,82 starburst as a more evolved episode.

More recently, the detection of SO$_2$,NS and NO emission in NGC\,253 have confirmed that shock
chemistry is at work in the nucleus of NGC\,253 (Mart\'{\i}n et al 2003, submitted).

\begin{figure}
\plotfiddle{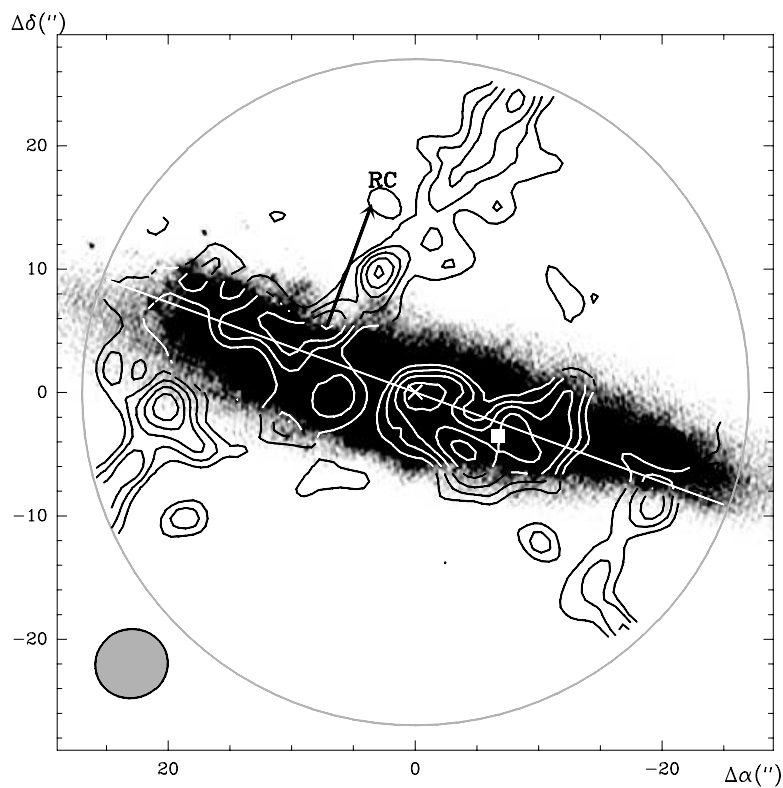}{9.15cm}{0}{60.}{60}{-180.}{-70.}
\caption{Integrated intensity map of SiO(v=0,J=2-1) emission (contours) in the central 700\,pc of
M\,82 obtained with the IRAM PdBI from Garc\'{\i}a-Burillo et al. (2001). The SiO map is overlaid
with the  4.8\,GHz--radio continuum image (grey scale) of Wills et al. (1999). Two major features
unveil large-scale shocks in the disk-halo interface of M\,82: a 500\,pc chimney (coincident with a
radio continuum chimney: RC) and a supershell enclosing SNR 41.95+57.5 (squared marker).}
\end{figure}

\begin{figure}
\plotone{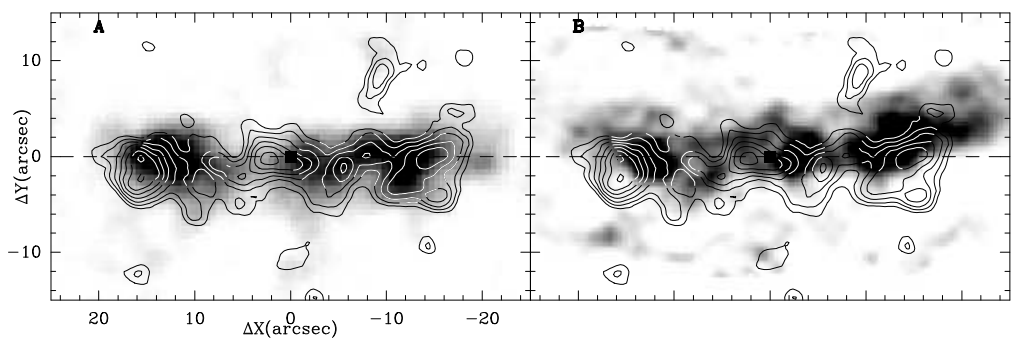}
\caption{Propagation of PDR chemistry in the M\,82 nuclear starburst is probed by the detection of
widespread HCO (F=2-1) emission (from Garc\'{\i}a-Burillo et al. 2002). The nested ring morphology
of the 650\,pc HCO disk (in contours) is compared with the H$^{13}$CO (A) and CO (B) disk emissions
(both in grey scale). The strong variations of the line intensity ratios indicate the propagation
of PDR chemistry in M\,82.} 
\end{figure}


\subsection{Propagation of PDR Chemistry in SB}

We have mapped the emission of the formyl radical (HCO) in the nucleus of M\,82 (Garc\'{\i}a-Burillo
et al. 2002). HCO is known to be enhanced in the interfaces between the ionized and molecular gas,
making it a privileged tracer of PDR. The 5'' HCO map of M\,82, the first obtained in an external
galaxy, shows a ring-like distribution, also displayed by other molecular/ionized gas tracers
in this galaxy. Most remarkably, the rings traced by HCO, CO and HII regions are nested, with the
HCO ring lying in the outer edge of the molecular torus (see Fig. 2).
The high overall abundance of HCO in M\,82 ($\sim$4$\times$10$^{-10}$) indicates that its nuclear
disk has become a giant PDR of $\sim$650\,pc size. Furthermore, the existence of a nested ring
pattern with the highest HCO abundance being estimated in the outer ring
($\sim$0.8$\times$10$^{-9}$), suggests that PDR chemistry is propagating in the disk.   

The PdBI maps of M\,82 made in SiO and HCO illustrate how two different gas chemistry scenarios can
be simultaneously at play in the same galaxy. The strong UV fields of the M\,82 starburst have
created a giant PDR inside the disk, while the expansion of hot gas created by successive SN
explosions entrains neutral gas into the halo driving shocks located in the disk-halo interface of
the galaxy.   
   
\subsection{XDR Chemistry in AGN: the nucleus of NGC\,1068}

Molecular gas in the CND of AGN hosts can be exposed to strong X-ray irradiation. Contrary to UV
photons, X-rays can penetrate gas column densities out to N(H$_2$)$\sim$10$^{23-24}$cm$^{-2}$
before being attenuated. Therefore, XDR could be the dominant sources of emission for the molecular
gas in the vicinity of AGN (Maloney et al. 1996). Tantalizing evidence that the chemistry of
molecular gas in the CND of AGN is `exotic' came from the large HCN/CO abundance ratio measured in
the nucleus of the Seyfert 2 galaxy NGC\,1068 (Tacconi et al. 1994), first interpreted as a
signature of enhanced oxygen depletion in the NLR. 

In the course of our ongoing 30m survey, we detected SiO emission in the starburst ring of
NGC\,1068. Most remarkably, we detected also SiO emission coming from the CND torus of NGC\,1068,
i.e., from a source mostly unrelated to recent star formation. We derived an SiO abundance enhanced
out to $\sim$10$^{-9}$ in the CND. Silicon chemistry in the CND of NGC\,1068 is driven either by
X-rays or by violent shocks near the central engine. To bring some light into the `obscuring torus
chemistry' we made complementary observations with the 30m telescope and PdBI for eight molecular
species, purposely chosen to fully explore the predictions of XDR models for the molecular gas
phase. Observations included several lines of CN, HCO, H$^{13}$CO$^{+}$, H$^{12}$CO$^{+}$,
HOC$^{+}$, HCN, CS and CO. A first analysis of this survey, presented by Usero et al. (2003), has
shown that the bulk of the molecular gas emission in the CND of NGC\,1068 can be interpreted as
coming from a giant XDR.
         
\section{Conclusions}

The advent of highly sensitive interferometers has made possible to study the complex molecular   
inventory of galaxies, going beyond the classical `CO maps'. In particular, the information provided
by molecular gas tracers of peculiar chemistry scenarios such as PDR, XDR, and Shock Chemistry is a
fundamental tool to explore the physical/chemical evolution of the ISM content in SB and AGN. In the
course of the combined IRAM 30m+PdBI survey we have studied a limited sample of SB and AGN. The
first findings have revealed already clear study cases of large-scale shocks at work (NGC\,253,
M\,82), propagation of PDR chemistry in SB (M\,82) or XDR chemistry driven by AGN (NGC\,1068).

\end{document}